\documentclass[a4paper,aps,prl,twocolumn,amsmath,amssymb,showpacs]{revtex4}

\usepackage{graphicx}
\usepackage{amsmath}
\usepackage{amssymb}
\usepackage{color}
\usepackage{ifpdf}
\usepackage{epsfig}
\usepackage{graphicx}

\begin{document}

\title{Morphogen Profiles Can Be Optimised to Buffer Against Noise}

\author{Timothy E. Saunders and Martin Howard}

\affiliation{Dept of Computational and Systems Biology, John Innes
  Centre, Norwich, NR4 7UH, United Kingdom}

\begin{abstract}
Morphogen profiles play a vital role in biology by specifying position
in embryonic development. However, the factors that influence the
shape of a morphogen profile remain poorly understood. Since
morphogens should provide precise positional information, one
significant factor is the robustness of the profile to noise.  
We compare three classes of morphogen profiles (linear, exponential, algebraic) 
to see which is most precise when subject to both external embryo-to-embryo 
fluctuations and internal fluctuations due to intrinsically random processes 
such as diffusion. We find that both the kinetic parameters and the overall 
gradient shape (e.g. exponential versus algebraic) can be optimised to 
generate maximally precise positional information.
\end{abstract}


\maketitle

Morphogens are signaling molecules which play a vital role in
biological development by inducing responses in a
concentration-dependent manner \citep{Lander2007}.  In the standard
model of morphogen gradients, morphogen proteins originate from a
localized source, diffuse and are degraded, setting up a concentration
gradient across the system. This gradient can control patterns of gene
expression, where, for example, a gene is switched on when the
concentration is above a certain fixed threshold, but is off
otherwise.  In this work, we focus on morphogen gradients specifying
boundaries of gene expression at fixed absolute distances from the
morphogen source, a scenario that is frequently realised in
developmental biology.

Developmental systems need to be robust to sources of noise in order
to generate precise patterns of gene expression. Here, we address a
simple question: in the presence of noise, which morphogen profile
is most precise in specifying the positions of gene
expression boundaries? In principle, any spatially non-uniform profile
could be used to position gene expression boundaries; our goal is to
understand which profiles might be preferred. Previous theoretical
approaches have predominantly examined robustness of morphogen
profiles to embryo-to-embryo fluctuations in the morphogen production
rate \citep{Eldar2003, Bollenbach2005, Umulis2008}, and analysis
suggests that algebraic profiles are most precise \citep{Eldar2003}.  However,
even in (hypothetical) embryos with no embryo-to-embryo fluctuations,
there would still be variation in positional information due 
to internal fluctuations in morphogen production, diffusion and degradation
\citep{Elowitz2002}.  Such fluctuations impose limits on the precision
of biochemical signaling \citep{Berg1977,Bialek2005,Tostevin2007}, and
could in principle alter the shape of the profile with the best
precision.  Internal fluctuations are particularly large if the
morphogen is, directly or indirectly, a transcription factor.  In that
case, the arrival of a morphogen molecule at the nanometer-scale DNA binding
sites on the target genes will be a rare, stochastic event
\citep{Tostevin2007, Gregor2007b}. Moreover, internal fluctuations in
the morphogen production rate may also play an important role
\cite{England2005}.

Recent experiments in the fruit fly {\it Drosophila melanogaster} have
quantitatively studied the morphogen proteins Bicoid (Bcd)
\citep{Houchmandzadeh2002,Gregor2007a, Gregor2007b}, Decapentaplegic
(Dpp) \citep{Kicheva2007} and Wingless \citep{Kicheva2007}. 
Interestingly, the observed profiles are exponentially 
decaying in all cases.  In order
to better understand this finding, we compare three classes of
morphogen profiles (linear, exponential, algebraic) to see which
is most precise when subjected to the combined effects of both external and internal
fluctuations.  We find that the kinetic parameters describing
morphogen profiles can be optimised to buffer against the combined
effects of both sources of noise.  By comparing optimised profiles, we
then see that the overall shape of the profile (e.g. exponential
versus algebraic) can also be optimised.  Exponential profiles
frequently emerge as the best compromise: such profiles are not
particularly robust to either external or internal fluctuations taken
singly, but when both types of fluctuation are taken together,
exponential profiles can be the most precise. We therefore propose a
simple design principle for morphogen profiles, namely that evolution
has selected gradients with optimal robustness to the combined effects
of embryo-to-embryo and internal noise.  Depending on which source of
noise is most important, qualitatively different morphogen profiles
will be selected. Given that very high positional precision can be
achieved by morphogens (e.g. a few percent of embryo length in the Bcd
system), it seems plausible that optimisation may well be exploited by
evolution. However, there may still be other constraints on morphogen systems,
for example, on the information capacity of the signaling network
\citep{Tkacik2008a,Emberly2008}.

{\it Models.} We consider three representative models of morphogen
gradients: a freely diffusing morphogen with a source and a sink
\citep{Crick1970}; diffusion with linear decay; and diffusion with
quadratic decay \citep{Eldar2003}.  Why study these three particular
models? Diffusion with linear decay leads to exponential morphogen
profiles that have been observed experimentally.  Non-linear decay
mechanisms leading to algebraic profiles are robust to external
morphogen production fluctuations \citep{Eldar2003} and we study
quadratic decay (decay via dimerisation) as a representative example
\citep{Eldar2003}.  Bone Morphogenetic Protein (BMP) in {\it Xenopus}
embryos is a possible candidate for a morphogen shaped by such
effective non-linear degradation \cite{Ben-Zvi2008}.  Finally, the
source-sink model generating a linear profile provides insight into
gradients generated by diffusion with localised degradation
\citep{Crick1970}.  This is realisable when the morphogen degradation
factors are tightly localised, such as for retinoic acid along the
anterio-posterior axis of mice embryos \cite{Sakai2001}.

A three-dimensional system is considered with a planar source of
morphogen at $x=0$ that produces a flux of proteins ($J$) which
diffuse ($D$) through the system (length $L$). In the absence of
fluctuations the morphogen concentration is given by the
reaction-diffusion equation
\begin{equation}
\label{eq:react-diff}
\partial_t \rho(\mathbf{x},t) = D \nabla^2 \rho(\mathbf{x},t) -
\psi\rho(\mathbf{x},t)^n \,,
\end{equation}
where $\psi=\mu[s^{-1}]$ ($n=1$) and $\psi=\alpha[\mu m^{3}s^{-1}]$
($n=2$). The boundary conditions are $D\partial_x\rho(x)|_{x=0}+J=0$
and, for $n=1,2$, $D\partial_x \rho(x)|_{x=L}=0$. The source-sink
model corresponds to $\psi=0$ and a sink at the boundary $x=L$
(i.e. $\rho(L)=0$). The steady-state solutions to
(\ref{eq:react-diff}) depend only on the linear distance from the
source in the $x$-direction:
\begin{eqnarray}
\label{eq:crick}
\rho_{s-s}(x) &=& (J/D)(L-x)\,, \\
\label{eq:linear}
\rho_{lin}(x) &=& (J\lambda/D)e^{-x/\lambda} \,, \\
\label{eq:non-linear}
\rho_{quad}(x) &=& A(x+x_0)^{-2} \,,
\end{eqnarray}
where $\rho_{s-s,lin,quad}$ correspond to the source-sink, linear
decay and quadratic decay models respectively and
$\lambda=\sqrt{D/\mu}$, $A =6D/\alpha$, $x_0 = (12D^2/J\alpha)^{1/3}$
with (\ref{eq:linear},\ref{eq:non-linear}) valid for $x,x_0,\lambda
\ll L$.

{\it External fluctuations.} We first investigate robustness to
external fluctuations solely in $J$ as this is believed to be
an experimentally relevant scenario \citep{Gregor2007b}. The position $x_T$ where
the concentration passes through a threshold ($\rho_T$) varies as a
result of embryo-to-embryo fluctuations (${\delta}J$) in the morphogen
production rate. Keeping the threshold concentration fixed, and
expanding around $x_T$ to leading order, the width due to external
fluctuations is given by $W\approx\Delta\rho(x_T)
/|\langle\rho'(x_T)\rangle|$.  Here, $\Delta \rho(x_T) = [\langle
  \rho(x_T)^2 \rangle -\langle \rho(x_T)\rangle^2]^{1/2}$
($\langle...\rangle$ denotes ensemble averaging) are the variations
due to external embryo-to-embryo fluctuations, and
$\rho'(x_T)=\partial_x\rho|_{x=x_T}$. The widths for the three models
are $W_{s-s} \approx (L-x_T)({\delta}J/J)$, $W_{lin} \approx
\lambda({\delta}J/J)$ and $W_{quad} \approx (1/3)x_0({\delta}J/J)$, to
leading order in ${\delta}J/J$. Typically, $x_T \sim L/2$ (as in the
Bcd controlled {\it hunchback} ({\it hb}) gene expression boundary in
{\it Drosophila}) and $x_0<3\lambda$, leading to $W_{quad} < W_{lin} < W_{s-s}$.

\begin{figure}
   \begin{center}
   \includegraphics*[width=3.0in]{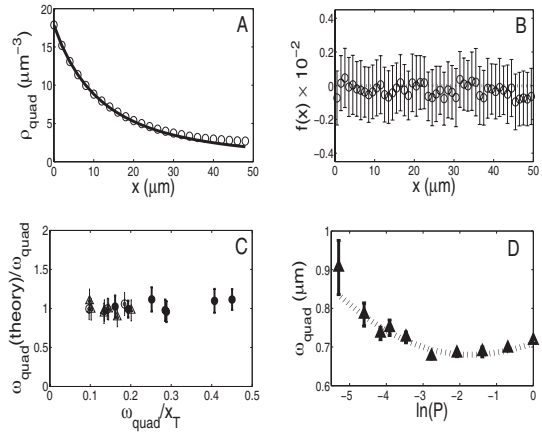}
      \caption{\label{fig1} Numerical Results. {\bf A}: $\rho_{quad}$
        against $x$, averaged for $\tau=5000s$. Line is fit to
        (\ref{eq:non-linear}). {\bf B}: Mean and standard deviation of
        $f(x)$, (error bars from 10 simulations).  {\bf C}: Comparing
        $\omega_{quad}$ from simulations with (\ref{eq:pow-width})
        for $J=1 (\bullet)$, $J=2 (\circ)$ and $J=5
        (\triangle)\,\textrm{molecules}\,{\mu}m^{-2}s^{-1} $, with
        various $\tau$ and $x_T$. {\bf D}: $\omega_{quad}$ against
        non-linear decay probability, averaged for $\tau=1000s$. Dashed line is
        prediction from (\ref{eq:pow-width}), where
        $\alpha=P\gamma/(1+cP\gamma)$ \citep{Saunders2009b,Tauber2005}
        ($\gamma={(\delta}x)^3/\delta{t}$, $c=12.5{\mu}m^{-3}s$).
        Parameters unless stated otherwise:
        $J=1\,\textrm{molecules}\,{\mu}m^{-2}s^{-1}$; $x_T=3{\mu}m$;
        $D=D_0=0.67{\mu}m^2s^{-1}$; $P=10^{-2}$; $L=50{\mu}m$; lattice
        spacing ${\delta}x=0.01{\mu}m$; time step $\delta t=2.5\times
        10^{-4} s$.}
   \end{center}
\end{figure}

{\it Internal fluctuations}. Within an embryo internal fluctuations
are also an important source of noise. We consider morphogen
production, diffusion and (if appropriate) degradation to be
stochastic processes. The morphogen concentration is sampled in a
region of linear size $\Delta x$ corresponding to the size of the
binding site at target genes. Note that incorporating
details of the binding process is unlikely to alter our results
\citep{Bialek2005}.  Internal fluctuations in particle density within
the sampling volume again cause the position where the gradient passes
through $\rho_T$ to vary, leading to imprecision in the positional
information provided by the gradient.  The width due to internal
fluctuations $\omega$ is again given by
$\omega\approx\Delta\rho(x_T)/|\langle\rho'(x_T)\rangle|$
\citep{Tostevin2007}, but where $\Delta\rho$ is now due to internal
fluctuations. In the source-sink and linear decay models the
statistics of the particle number $n(x)$ due to internal fluctuations
are Poissonian, since both models are linear and morphogen production,
diffusion and degradation (if present) are all Poisson processes
\citep{Tostevin2007,Wu2007,Lepzelter2008}.  We present numerical
results below demonstrating that the particle number statistics in the
quadratic decay model are also effectively described by a Poisson
distribution. This result is not obvious, as non-linear decay
processes could, in principle, generate non-Poisson fluctuations.  In
all cases, we assume that diffusion of morphogen proteins is a purely
three dimensional process (there could be one dimensional sliding
along DNA but this leads to similar fluctuations as three dimensional
diffusion \citep{Tkacik2007}).

{\it Simulations}. Stochastic simulations were performed on a
three-dimensional lattice containing discrete particles with particle
injection on the plane $x=0$, diffusion and (if appropriate)
degradation.  For the quadratic decay model, simulations were
performed using a range of parameter values for $J$, $D$, $P$
(probability of two particles degrading within a single time step
given that they occupy the same lattice site) and system size L.  In
Fig.~\ref{fig1}A we demonstrate that concentrations measured in our
simulations agree well with (\ref{eq:non-linear}) (the finite size
effects at large $x$ do not alter our conclusions
\cite{Saunders2009b}).  To confirm that particle number fluctuations
are described by Poisson statistics for the quadratic decay model we
calculated $f(x) = [\langle n(x)^2 \rangle - \langle n(x)
  \rangle^2-\langle n(x)\rangle]/\langle n(x) \rangle$, where each
simulation was averaged for 5000s (Fig.~\ref{fig1}B).  In three
dimensions diffusion is efficient enough to prevent the build-up of
non-Poissonian correlations resulting from the quadratic decay
reactions. This result is robust to parameter variations in our
simulations (data not shown). 

{\it Time/spatial averaging.} We now consider the effects of time and
spatial averaging \citep{Tostevin2007,Gregor2007b}, which act to
reduce $\omega$. Time-averaging is performed by the down-stream
signal-processing network, where the timescale is typically given by
the lifetimes of the transcripts/proteins of the target gene. Over an
averaging period $\tau$, there can, at most, be
$N_{\tau}=\tau/\tau_{ind}$ independent readings of the morphogen
concentration which reduce the measurement width by a factor $\sim
N_{\tau}^{-1/2}$. Intuitively, $\tau_{ind}\sim (\Delta{x})^2/D_0$, the
typical timescale for the sampling volume to empty and then be
refilled by new protein particles via diffusion \citep{Tostevin2007,
  Berg1977} ($D_0$ is `local' diffusion responsible for movement into
the sampling volume, which may not equal $D$, the bulk 
diffusion).  A constant associated with time averaging is found
numerically (see below). We also include spatial averaging, motivated
by recent experiments on the Bcd-{\it hb} signaling pathway
\citep{Gregor2007b}.  By averaging over a number of different
nuclei/cells, $N_{spat}$, the effects of internal fluctuations can be
further reduced by a factor $\sim N^{-1/2}_{spat}$ (up to a constant
of $\mathcal{O}(1)$, which we have verified numerically does not alter
our conclusions).  From \citep{Gregor2007b}, $N_{spat}\approx
CD_0\tau$ where $C$ is a constant which depends on the particular
arrangement of nuclei/cells.

{\it Width due to internal fluctuations.} For the three models we find
that the widths $\omega$ due to internal fluctuations are:
\begin{eqnarray}
\label{eq:sink-width}
\omega^2_{s-s} &=& \frac{k^2_0(D/D_0)}{{\tau}J\Delta xN_{spat}}
(L-x_T)\,,\\
\label{eq:exp-width}
\omega^2_{lin} &=& \frac{k^2_1(D/D_0)}{{\tau}J\Delta xN_{spat}}
        {\lambda}e^{x_T/\lambda}\,,\\
\label{eq:pow-width}
\omega^2_{quad} &=& \frac{k^2_2(D/D_0)}{2{\tau}J\Delta xN_{spat}}
\frac{(x_T+x_0)^4}{x^3_0}\,.
\end{eqnarray}
In Fig.~\ref{fig1}C, we compare (\ref{eq:pow-width}) with simulation
results for a range of parameter values. We find good agreement between
the two approaches, where $k_2=0.56\pm0.06$ is a fitting parameter. A
similar procedure for the other two models yields $k_0=0.53\pm0.07$
and $k_1=0.60\pm0.05$.

{\it Optimizing kinetic parameters.} Importantly, the
underlying kinetic parameters in the linear and quadratic decay
models can be optimised to minimise $\omega_{lin}$ and $\omega_{quad}$
at the threshold position $x_T$. The existence of optimal decay rates
is a general feature of morphogen gradients and occurs for all decay
exponents $n>0$ \cite{Saunders2009b}. Maximum precision is achieved
when $\lambda=x_T$ \citep{Tostevin2007} and $x_0=3x_T$. For the
non-linear decay model we verified numerically the existence of such
an optimal decay rate, see Fig.~1D.  In general, in the experimentally
relevant range $x_T \sim L/2$, $x_0\sim\lambda$, we find $\omega_{s-s} < \omega_{lin} <
\omega_{quad}$. Although both the morphogen density and slope affect
$\omega$, the value of the slope is the more important: the
source-sink model has the steepest, and the quadratic model the least
steep, profile, thereby generating the above ordering.  Comparing this
inequality with our earlier result $W_{quad}<W_{lin}<W_{s-s}$, the
ordering is reversed, so that, for example, the quadratic decay model
performs best on external fluctuations but worst on internal
fluctuations.

\begin{figure}
   \begin{center}
     \includegraphics*[width=3.0in]{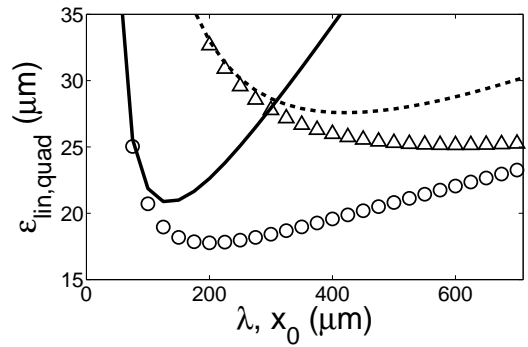}
      \caption{\label{fig2} Optimising morphogen profiles.
        ${\epsilon}_{lin}$ versus $\lambda$ (solid line) and
        ${\epsilon}_{quad}$ versus $x_0$ (dashed line) with
        $\omega_{lin} ({\circ})$ and $\omega_{quad} (\triangle)$.
        $J=1.0\,{\textrm{molecules}}\,{\mu}m^{-2}s^{-1}$,
        ${\delta}J/J=0.07$, ${\Delta}x=3\times10^{-3}{\mu}m$,
        $D=15{\mu}m^2s^{-1}$, $D_0=0.5{\mu}m^2s^{-1}$,
        $x_T=200{\mu}m$, $\tau=5$ minutes, $N_{spat}=18$. The
        contributions solely from external fluctuations, which are
        linear in $\lambda$, $x_0$ respectively, are omitted for
        clarity.}
   \end{center}
\end{figure}

 The robustness of our three models to the combined effects of internal
and external fluctuations can now be compared.  Since internal and
external fluctuations are statistically independent, the total width
is given by $\epsilon= \sqrt{\omega^2 + W^2}$.  We find that the
minimum in $\epsilon_{lin}$ with respect to $\lambda$ (and in
$\epsilon_{quad}$ with respect to $x_0$) becomes more pronounced than
is the case with internal noise alone (see Fig.~2). As a result, the
penalty in terms of reduced precision when using non-optimized
parameters is now more severe than in the internal noise-only case.
Note that the optimal values of $\lambda$ and $x_0$ are reduced
compared to their values when considering internal noise alone, see
Fig.~\ref{fig2}.

\begin{figure}
   \begin{center}
		\includegraphics*[width=3.3in]{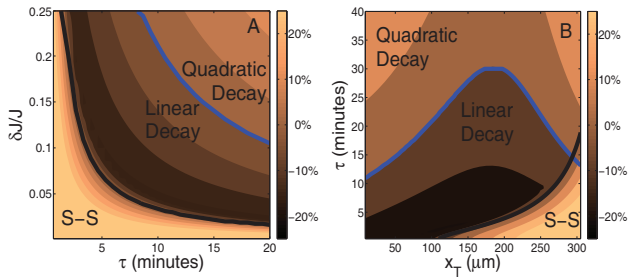}
      \caption{\label{fig3} {\bf A}: Effective phase diagram for 
      	relative position of the
        three models considered with varying internal and external
        fluctuations. {\bf B}: Relative precision of models with varying $x_T$
        in presence of internal fluctuations, with fixed
        ${\delta}J/J=0.07$.  Colourmap defined in text, $L=500{\mu}m$
        and other parameters as in Fig.~\ref{fig2}. Solid lines correspond to
        $\epsilon_{lin}=\epsilon_{s-s}$ (black) and
        $\epsilon_{lin}=\epsilon_{quad}$ (blue).}
   \end{center}
\end{figure}

{\it Optimizing morphogen profile shape.} We investigate which
model gives the most precise positional information when subjected to
embryo-to-embryo fluctuations (parameterised by ${\delta}J/J$) and
internal fluctuations (parameterised by the averaging time, $\tau$).
For each ${\delta}J/J$ and $\tau$, we compute the optimal decay
rates for the linear and quadratic decay models as above.
We then build up an effective phase diagram in the
$\tau-{\delta}J/J$ plane, determining the most precise
morphogen profile for particular levels of external and internal
noise, see Fig.~\ref{fig3}A. The parameters values used 
are similar to those found from experiments on {\it
  hb} expression in the {\it Drosophila} embryo
\citep{Houchmandzadeh2002,Gregor2005,Gregor2007a,Gregor2007b}. 
$J$ is kept constant between the models, representing fixed resource
expenditure.  The colourmap is defined as
$(\epsilon_{lin}-\epsilon_a)/\epsilon_{lin}\times100\%$ where
$\epsilon_a$ is the smaller of $\epsilon_{quad}$ and $\epsilon_{s-s}$.
The percentage change is negative when the linear decay profile is
most precise.  For small times or small $\delta J/J$, profiles from
the sink-source model are preferred as this model is best able to
buffer the dominant internal fluctuations. For large $\delta J/J$ or
large $\tau$ quadratic decay profiles are selected, as now the
quadratic model is now able to best buffer the dominant external
fluctuations. At intermediate values of ${\delta}J/J$ and $\tau$, when
$W$ and $\omega$ are similar, exponential profiles are preferred as
they provide the best compromise between robustness to both internal
and external noise.  From such diagrams we can build up a qualitative
understanding of why particular morphogen profiles are selected
depending upon the dominant sources of noise. We emphasise that our
methodology is general and not dependent on the specific parameter
values used above. For example, if $D=D_0$ then quadratic decay would
be favoured in a wider region of the phase space, but there would
still be conditions when linear decay, or the source-sink model, would
deliver higher precision. If we choose a general decay exponent $n$,
then, for a fixed $\delta J/J$, we find a qualitatively similar picture
with small $n$ favoured at short averaging times, with larger $n$
selected at longer times \cite{Saunders2009b}.

{\it Multiple targets.} In many developmental systems morphogens are
directly involved in the expression of several target genes at
different positions throughout the system (e.g. {\it orthodenticle}
($x_T\approx125{\mu}m$) \cite{Finkelstein1990} and {\it hb}
($x_T\approx240{\mu}m$) are both targets of Bcd). We consider
optimising the morphogen profiles at one expression boundary and then
compare how precisely the profiles determine a second boundary at a different
position (keeping ${\delta}J/J$ fixed).  Using the same optimised
morphogen profiles as above we find that exponential profiles, unlike
the other gradients, lead to the best precision at both
short and long distances for short times ($\tau<15$mins), Fig.~\ref{fig3}B. This
flexibility is a potential explanation for the widespread use of
exponential profiles in developmental systems.

We have applied our analysis to simple (though still
experimentally motivated) models of gradient formation, yielding
linear, exponential and algebraic profiles. Intriguingly, the best
characterized morphogens Bcd, Dpp and Wingless in {\it Drosophila} all
have exponential profiles and their decay lengths are significantly
less than their respective values of $x_T$
\citep{Houchmandzadeh2002,Kicheva2007}.  This is consistent with our
conclusion that decay lengths adapt to buffer the combined effects of
internal and external fluctuations and that exponential profiles
perform well when buffering combined internal-external fluctuations for
a range of $x_T$ (since all three morphogens have multiple target
genes). However, more complex processes may also be involved in the
formation of these gradients, including, for example, transcytosis, 
pre-steady-state measurement or mRNA gradients
\citep{Bollenbach2005, Bergmann2007, Spirov2009}. In some circumstances, morphogen systems
are able to scale with embryo length
\citep{Houchmandzadeh2002,Gregor2007b,He2008}. Nevertheless, once
these additional effects are better characterized, our concepts can
still be readily applied. Attaining maximal robustness to the
combined effects of internal and external noise may be a powerful
unifying principle in understanding the fundamental design of
morphogen systems.

We thank Enrico Coen and Richard Morris for insightful comments.
MH thanks The Royal Society for funding.

\end{document}